# Over-the-air Signaling in Cellular Networks: An Overview

Chunliang Yang

*Abstract*—To improve the capacity and coverage of current cellular networks, many advanced technologies such as massive MIMO, inter-cell coordination, small cells, device-to-device communications, and so on, are under studying. Many proposed techniques have been shown to offer significant performance improvement. Thus, the enabler of those techniques is of great importance. That is the necessary signaling which guarantee the operation of those techniques. The design and transmission of those signaling, especially the over-the-air (OTA) signaling, is challenging. In this article, we provide an overview of the OTA signaling in cellular networks to provide insights on the design of OTA signaling. Specifically, we first give a brief introduction of the OTA signaling in long term evolution (LTE), and then we discuss the challenges and requirements in designing the OTA signaling in cellular networks in detail. To better understand the OTA signaling, we give two important classifications of OTA signaling and address their properties and applications. Finally, we propose a signature-based signaling named (single-tone signaling, STS) which can be used for inter-cell OTA signaling and is especially useful and robust in multi-signal scenario. Simulation results are given to compare the detection performance of different OTA signaling.

## I. INTRODUCTION

With the widespread of various smart devices, e.g., smartphones, and the introduction of new infrastructures, the cellular network becomes more and more complicated. Both the industry and the academia are facing never-seen challenges in managing the whole cellular network. Among those challenges, the first one is how to increase the capacity to meet the rapid growth of data traffic; the second challenge is how to design various reliable signaling in cellular networks without causing too much overhead (e.g., bandwidth and power); the third challenge is how to extent the coverage of the current cellular network to give user a flat/smooth experience of service; the fourth challenge is how to reduce the delay to provide high quality of real-time service. Each of those challenges has attracted a lot of interests. In this article, we focus the discussion on the second challenge, i.e., the signaling design, especially the over-the-air (OTA) signaling design.

The various signaling in cellular networks are of special importance and a must since they guarantee the network to operate in an efficient, stable and reliable way. They may either carry system parameters (e.g., FFT size or transmission bandwidth, cell identity (ID), cyclic prefix length, etc.) for configuring mobile devices, or carry information (e.g., transmission frame format, resource allocation, power control command, modulation and coding scheme, new data indicator, redundancy version, precoding information, channel quality indicator feedback, hybrid ARQ indicator, etc.) for controlling the transmission of data traffic. Even a simple procedure may need many signaling messages. For example, to establish a connection from a mobile device to a base station, up to 30 signaling messages are needed in a third-generation (3G) cellular network and 7~13 signaling messages are needed in LTE [1]. Since the transmission of data traffic is controlled by signaling, it will fail if the signaling transmission is in error.

As more and more advanced technologies are employed in cellular networks, the need for corresponding signaling is explosive. For example, the coordination of neighboring cells [2], the introduction of small cells into the current macro cellular network and the integration of Wi-Fi network and cellular network [3], the support of peer-to-peer or device-to-device communications in cellular networks [4]-[6], the use of massive MIMO [7], [8], the mobility management in heterogeneous networks [9], , all these new technologies need new signaling to facilitate their operation. Study showed that the signaling traffic increased more than two hundred percent over a two-year-period at a sample European 3G operator [1]. And this is the motivation of this article, i.e., to overview the current OTA signaling and to provide some insights on the design of future OTA signaling.

Since the signaling is a kind of overhead, more signaling means more overhead. Thus, the transmission of signaling should use as less resource as possible while provide accurate and reliable control information.

This article gives an overview of the OTA signaling in cellular networks. Section II gives a brief introduction of various OTA signaling in cellular networks. Section III describes the challenges and requirements in designing OTA signaling. Section IV gives two typical classifications of OTA signaling. A new signature-based signaling scheme is proposed in Section V. Section VI compares the detection performance of different signaling schemes. Section VII concludes this article.

## II. THE OTA SIGNALING IN CELLULAR NETWORKS

As discussed in last section, the OTA signaling enables the operation of the whole cellular network, e.g., configuring the mobile devices, communicating the status of certain entities in the network and controlling the behavior of certain entities. Thus, in this section, we give a brief introduction of various necessary OTA signaling in cellular networks taking the LTE standard as an example. In LTE, the downlink OTA signaling includes the synchronization signaling, the physical broadcast signaling, the physical control format indicator signaling, the physical downlink control signaling, the physical hybrid automatic repeat request (ARQ) indicator signaling and so on; the uplink signaling includes the random access signaling, the physical uplink control signaling, and so on.

Just as their names implied, the synchronization signals including the primary synchronization signal (PSS) and the secondary synchronization signal (SSS) enable the acquisition of the downlink received symbol timing and frequency. The cell identity (ID) information is also carried on the synchronization signals. The physical broadcast signaling transmitted in the physical broadcast channel (PBCH) includes a certain set of cell- and/or system-specific information to the mobile devices (or user equipments, UEs). After receiving the synchronization signals and the physical broadcast signaling, the mobile devices generally acquire information that includes the overall transmission bandwidth of the cell, the cell ID, the number of transmit antenna ports and the cyclic prefix length, etc. These OTA signaling is designed for the scenario where a mobile device is turned on and the scenario where the mobile device loses the downlink time and frequency synchronization.

Before the mobile device receives and decodes the downlink data traffic, it has to decode three downlink control signaling which are transmitted every subframe, i.e., they are physical control format indicator signaling transmitted in physical control format indicator channel (PCFICH), physical downlink control signaling transmitted in physical downlink control channel (PDCCH) and physical hybrid ARQ indicator signaling transmitted in physical hybrid ARQ indicator channel (PHICH). The PCFICH carries information of the number of Orthogonal Frequency Division Multiplexing (OFDM) symbols used for PDCCH in this subframe. The PDCCH is used to inform the mobile device about the downlink/uplink scheduling assignments, e.g., the resource allocation as well as modulation and coding scheme and hybrid ARQ control information. The PHICH is used to carry hybrid ARQ ACK/NACK feedback for uplink traffic data transmissions.

On the uplink, the random access signaling transmitted in random access channel (RACH) is used by mobile devices for random access. Random access is generally performed when the mobile device turns on from sleep mode, performs handoff from one cell to another or when it loses uplink timing synchronization. Random access allows the base station (or Evolved Node B, eNB) to estimate and, if needed, adjust the mobile device uplink transmission timing to within a fraction of the cyclic prefix.

The uplink feedback control signaling carries the information of channel quality, MIMO channel rank, preferred precoding matrix and hybrid ARQ ACK/NACK. The channel quality information is used for channel sensitive scheduling as well as link adaptation in the downlink. The MIMO rank and precoding information is used to select a suitable MIMO transmission format which includes the number of MIMO layers transmitted and antenna weights for beamforming. The hybrid ARQ ACK/NACK is used to support downlink hybrid ARQ operation. The periodic feedback (channel quality, MIMO channel rank, preferred precoding matrix) and hybrid ARQ ACK/NACK are transmitted on physical uplink control channel (PUCCH). The aperiodic feedback of channel quality, MIMO channel rank and preferred precoding matrix is performed on physical uplink shared channel (PUSCH).

As we can see, the aforementioned various OTA signaling deliver a lot of special and necessary control information for the transmission of data traffic in the whole cellular network. Their design goals and requirements are different from that of traffic data. Thus, it's interesting to know those unique design goals and requirements.

## III. CHALLENGES AND REQUIREMENTS IN DESIGNING OTA SIGNALING

Compared to wired signaling design, OTA signaling design is more challenging since OTA signaling is transmitted over the wireless channel which not only has time-varying and random large scale fading (i.e., pathloss) and small scale fading (multipath and Rayleigh fading) but also has limited bandwidth. Besides, the transmission of OTA signaling may be interfered by the transmission of data traffic as well as many other OTA signaling. Thus, at the receiver, the received OTA signaling signal may be severely corrupted by the wireless channel, impaired by the noise and interfered by other signals.

Due to the complex transmission environment and the special functions of OTA signaling, e.g., configuring the receiver, and controlling the transmission and reception of data traffic, the design of OTA signaling has its own targets and requirements. Firstly, high reliability and robustness of OTA signaling must be achieved since the data traffic transmission would fail when OTA signaling transmission fails. Secondly, unlike the transmission of data traffic in which hybrid ARQ mechanism is employed to retransmit the same data when transmission is failed, the same OTA signaling is transmitted only once to reduce overhead since the transmission of OTA signaling needs precious wireless resources. Moreover, during the one-shot transmission, the OTA signaling should use as less resources (e.g., frequency, time, power) as possible to reduce overhead. Thirdly, since the OTA signaling would encounter large scale pathloss attenuation in the wireless channel, large coverage of OTA signaling is usually needed to guarantee their reliable reception for mobile devices locating at different places. Finally, the processing of OTA signaling cannot be too complex since too complex processing would lead to delayed OTA signaling which affects the transmission or reception of data traffic. Therefore, the main design goals and requirements of OTA signaling are high reliability and robustness, low overhead, large coverage, low processing complexity and low delay.

### A. High Reliability

The transmission scheme of OTA signaling may differ from that of data traffic. As the signaling enables the reception of the actual data traffic, it should naturally be more robust and reliable against channel impairments than data traffic itself. To improve the reliability and robustness, the first idea is to let OTA signaling transmit in a dedicated channel. That is to dedicatedly allocate a part of the total system resource (e.g., frequency, time, or code) to the transmission of OTA signaling. By doing so, the transmission of OTA signaling will not be interfered by data traffic since data traffic are not allowed to transmit in those dedicated channels. In current cellular communication systems, almost all the OTA signaling is transmitted in dedicated channels to avoid the interference from data traffic. For example, in LTE, the PSS and SSS are carried in the last and second last OFDM symbols respectively in slot number 0 and slot number 10, and they are carried in the frequency domain over the middle six resource blocks using 62 subcarriers out of a total of 72 subcarriers (1.08 MHz); the physical broadcast signaling is transmitted on the first four OFDM symbols of the second slot of the first subframe within a radio frame occupying the central six resource blocks; the downlink control information including control format indicator (CFI), uplink and downlink resource assignments, and the uplink hybrid ARQ ACK/NACK is always carried on the first three or four OFDM symbols of a subframe. On these resources, data traffic is not allowed to be transmitted on.

However, despite of the interference from data traffic, the OTA signaling may interfere with each other since multiple OTA signaling may simultaneously transmit on the same dedicated channel. Therefore, to further improve the reliability and robustness of OTA signaling, many advanced techniques, e.g., low code rate coding, interleaving, spreading, scrambling, exploit diversity (e.g., time, frequency, space) and so on, can be employed. In the downlink of LTE, the control format indicator message is coded using a (3,2) simplex code with 10 repetitions to guarantee its reliable transmission [10]; for the transmission of PDCCH, cell-specific interleaving was employed to reduce interference; the hybrid ARQ indicator transmitted in the PHICH is spread using orthogonal sequences to reduce interference; meanwhile, all the transmissions of CFI, PDCCH and hybrid ARQ indicator are using transmit diversity and scrambling to improve reliability and reduce interference. In the uplink, either in a CDMA system (e.g., WCDMA, 1xEV-DO [11], [12]) or an orthogonal frequency division multiple access (OFDMA) system (e.g., LTE [10]), random access signals transmitted in the uplink random access channel use direct-sequence spreading techniques or its variants for random access to reduce inter-signal interference. Pseudo-Noise (PN) based sequences are used in WCDMA and 1xEV-DO. In LTE, prime-length Zadoff-Chu sequences have been chosen for improved orthogonality between RACH signals.

Despite high reliability requirement of OTA signaling compared to data traffic, different OTA signaling has different error requirements. For example, in the uplink of LTE, the hybrid ARQ ACK/NACK feedback has strictly low error rate requirement than the channel quality information.

*B. Low Overhead*

Low overhead is another important design goal of OTA signaling. However, OTA signaling does not come for free. It requires considerable precious wireless resources to guarantee its reliable transmission. If many resources are used by OTA signaling, then no resources are left for data traffic transmission. As a matter of fact, low overhead and high robustness / reliability contradict with each other to some extent. The overhead increases while employing many advanced techniques to improve reliability and robustness. For example, low code rate coding using more resources than high code rate coding for the same number of information bits. The overhead may not be a problem in current macrocell networks since most existing OTA signaling schemes are designed only for communicating to the (nearest) *serving* base stations which may not need too many resources. However, as the cellular becomes more and more complicated, e.g., the introduction of small cell into the current macro cellular network, the integration of Wi-Fi network and cellular network, the support of peer-to-peer or device-to-device communications in cellular networks, the mobility management between different generations of cellular networks, e.g., GSM and WCDMA, 1xEV-DO and LTE, and so on, more OTA signaling are needed to facilitate the operation of those new techniques. For example, as small cells are introduced in the current macrocell networks, there is high possibility that inter-cell signaling are needed for various purposes (e.g., inter-cell coordination to reduce interference and improve capacity) due to various reasons (e.g., slow and limited backhaul). Thus, this increased demand of signaling strongly challenges the design of low overhead OTA signaling.

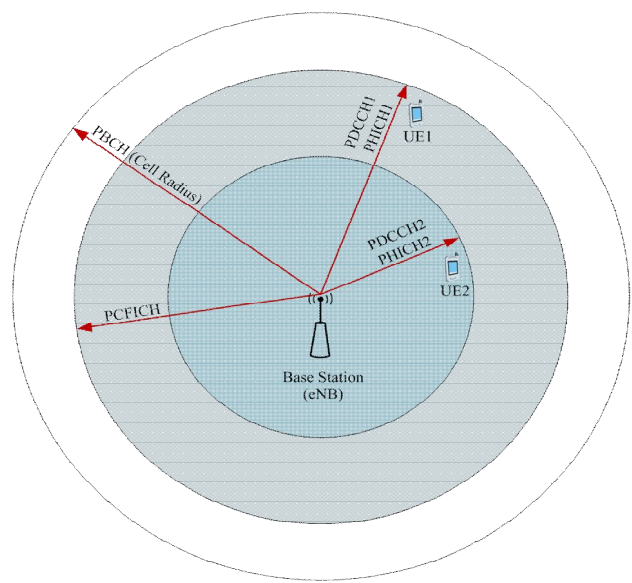

Figure 1. Illustration of coverage of different OTA signaling in a cellular network.

*C. Large Coverage*

Since the OTA signaling serves the traffic data transmission, its coverage must be larger than the corresponding data traffic. Different OTA signaling carry different information, thus they have different functions and target receivers, therefore, different coverage. For broadcast signaling (targeted for all the mobile devices in the cell), its coverage must reach the cell edge since there may be mobile devices locating at cell edge. For the OTA signaling that is targeted for only one or a group of mobile devices, its coverage must be reach to the mobile device that has the worst channel condition. Let's take the downlink control channel in LTE as an example shown in Figure 1.

In Figure 1, the PBCH (targeted for all the UEs) covers the whole cell. To guarantee the reception of PCFICH for the scheduled UE with the worst channel condition in the current subframe, the PCFICH must cover UE1 (assume UE1 and UE2 are scheduled and the channel condition of UE2 is better than that of UE1 in the current subframe). The PHICH and PDCCH just need to cover the corresponding UE via power control. As discussed in last section, with the development of cellular networks, inter-cell OTA signaling for coordination may be needed in the future which requires deep coverage as the target receivers for the inter-cell OTA signaling are the *neighboring* base stations instead of the *serving* base stations.

*D. Low Complexity and Delay*

Since the processing of OTA signaling is ahead of the processing of data traffic and is the first step, it should need as less priori system information or parameters as possible. For example, the processing of PSS, SSS, and PBCH does not need system bandwidth information in LTE. Besides, the processing of OTA signaling should be as simple as possible. If it is too complex, not only the power is wasted and the situation is even worse when the receiver is a mobile device which use a battery, but also the delay of OTA signaling is introduced which may have negative effect or even damage on the data traffic transmission and reception. However, as discussed above, OTA signaling also needs high reliability and robustness. Thus, high reliability and robustness sometimes means high complexity. Therefore, tradeoff must be made between high reliability / robustness and low complexity and delay.

Generally speaking, considering the unique function of OTA signaling, the design rule of OTA signaling is to use as less resource as possible to reduce overhead and improve spectral efficiency, and to use as simple scheme and processing as possible to reduce complexity and delay, while at the same time to guarantee high reliability and robustness within certain coverage area.

## IV. CLASSIFICATION OF OTA SIGNALING

In this section, we provide two typical classifications of OTA signaling.

*A. Packet- and Signature-based Signaling*

Although the OTA signaling carry many different types of information, they can be classified as packet-based signaling and signature-based signaling according to their way of conveying information.

*1) Packet-based Signaling*

Packet-based signaling uses an encoded packet that includes a large amount of information, e.g., the frame format of data traffic, the resource allocation, precoding mode and so on. For example, in LTE the PDCCH packet includes the information of resource allocation, modulation and coding scheme, redundancy version, transmit power control command for scheduled PUSCH, CQI request, precoding information and so on. This kind of signaling is processed in the same way as the regular data traffic. The advantage of the packet-based signaling scheme is that much information related to the processing of corresponding data traffic can be obtained from the signaling message by one transmission. However, the drawback is that the packet is large in size, and hence has to be heavily channel-coded for spectral efficiency, using, e.g., convolutional, LDPC, or turbo codes. Thus the signaling information has to go through a series of complex processing at the transmitter and receiver. For example, the information contained in the PDCCH is UE-specific CRC attached, tail-biting convolutionally coded, rate matched, and QPSK-modulated just like a regular data traffic packet except that a PDCCH packet is transmitted at a higher power than the regular data traffic packet to ensure coverage. The transmission chain processing of PDCCH is shown in Figure 2. The resultant PDCCH packet is therefore complex in the sense that it requires pilots (for demodulation), CRC coding, channel coding, QPSK modulation, rate matching at the transmitter, and channel estimation and decoding employing, e.g., a Viterbi decoder at the receiver while maintaining the time and frequency synchronization. The resulting "heavy-weight" signaling message not only requires excessive resources and processing power but also is sensitive to any type of co-channel transmissions (collisions) which introduces interference and packet decoding failure. Thus it is transmitted in a dedicated channel. Moreover, such packet-based signaling is not power and resource efficient. A receiver has to employ high-complexity channel estimation and decoding to obtain the signaling packets from the transmitter after each scheduling.

Another advantage of packet-based signaling is that it can fully exploit time-frequency diversity since they employ channel coding. Thus the transmission is distributed over a large time duration and bandwidth to capture time-frequency diversity. For example, in LTE, the control format indicator (CFI) message is coded using a (3, 2) simplex code with 10 repetitions and is almost uniformly distributed in the frequency [10].

In LTE, the signaling transmitted in PBCH, PCFICH, PDCCH and PHICH all use packet-based signaling.

*2) Signature-based Signaling*

OTA signaling using a signature typically involves the transmission of a special simple physical layer waveform as the signaling signal is used to conveying simple and a small amount of control information (less than 10 bits). This kind of signaling is constructed from a set of sequences. The signature of each sequence represents information. Different sequences represent different information. For example, in LTE three

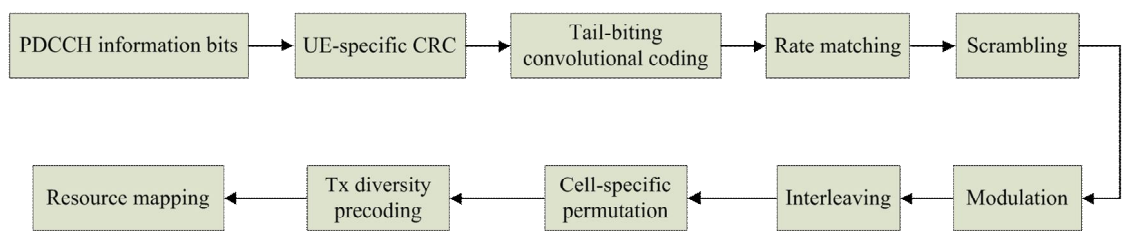

Figure 2. Transmission chain processing of PDCCH in LTE.

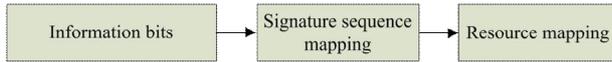

Figure 3. Transmission chain processing of signature-based signaling.

length-62 Zadoff-Chu sequences are used to conveying three physical layer cell identities. The transmission chain processing of signature-based signaling is shown in Figure 3. Compare Figure 2 and Figure 3, the processing of signature-based signaling is much simpler than that of packet-based signaling.

The advantage of the signature-based signaling is that the information contained in the signature is typically much less than that in a packet-based signaling therefore can be designed with higher detection performance and spectral efficiency, negating the need for pilots and other complex operations like channel coding and decoding.

Since correlation is performed to detect the signature-based signaling at the receiver, the orthogonality between sequences is one of the most important criteria for choosing sequences. Thus, sequences with good autocorrelation and cross-correlation are preferred. If perfect correlation (i.e., orthogonal) can be obtained, multiple signature sequences can be transmitted in the same channel without interfering with each other.

Despite explicitly carrying signaling information, signature-based signaling can also be used for time and frequency synchronization at the receiver, like the PSS and SSS in LTE.

One drawback of signature-based signaling is that it can only transmitted in continuous / consecutive time-frequency resource where the channel is nearly static to keep the good correlation property of sequences. For example, the perfect orthogonality of Zadoff-Chu sequences with different cyclic shift of the same root sequence is lost when the sequences undergo time and frequency selective fading.

Commonly used waveforms for signatures include PN, Walsh or Zadoff-Chu sequence modulated QPSK waveforms. They are currently used for synchronization and random access in cellular networks. For example in LTE, the Zadoff-Chu sequence is used to construct the PSS and the RACH, and the PN sequence is used to construct the SSS.

3) *Comparison between Packet- and Signature-based Signaling*

According to the discussion above, the packet-based signaling and the signature-based signaling have different properties and applications. Therefore, we will make a comparison of them in terms of payload size, detection, complexity, time and frequency synchronization.

**Payload size** The packet-based signaling scheme can carry a large amount of information bits, e.g., tens to thousands of bits, while the signature-based signaling scheme can only carry no more than ten bits. The reason is that the number of signature sequences determines how many information bits can be carried in a signature-based signaling scheme, and the number of signature sequences with good correlation properties is limited, especially the orthogonal sequences. Thus the number of information bits in signature-based signaling scheme cannot be very large.

TABLE I
COMPARISON OF SIGNATURE-BASED SIGNALING AND PACKET-BASED SIGNALING

| Type | Signature-based signaling | Packet-based signaling |
|---|---|---|
| Payload size | Small (< 10 bits) | Large (10s to 1000s bits) |
| Detection | Simple | Complex |
| Complexity | Small | Large |
| Time and frequency syn | Not needed | Needed |

**Detection** In the packet-based signaling scheme, the receiver needs to estimate the channel to decode the received packet to get the original information, which is a single signal detection and decoding. However, the detection of the signature-based signaling is similar to multiuser detection but is more complex. In the signature-based signaling scheme, what the receiver does is to determine the unknown set of signals sent from unknown different transmitters, while multiuser detection is to decode transmitted data streams from known transmitters / users. Hypothesis detection is usually needed in signature-based signaling scheme.

**Complexity** From Figure 2 and Figure 3, we can clearly see that the process of packet-based signaling scheme is more complex than that of the signature-based signaling scheme.

**Time and Frequency Synchronization** The signature-based signaling can not only carry a small amount of information bits, but also be used for time and frequency synchronization. But the packet-based signaling can only work after time and frequency synchronization have been accomplished.

Therefore, signature-based signaling scheme is used in scenarios where a small amount of information bits need to be transmitted, no synchronization is achieved and the receiver has to detect unknown set of signals. And the packet-based signaling scheme can be used in the scenarios where synchronization is achieved, a large amount of information bits need to be transmitted, and the receiver has strong processing power.

*B. Intra- and Inter-cell Signaling*

The OTA signaling can also be classified into intra-cell and inter-cell signaling according to their coverage.

*1) Intra-cell OTA Signaling*

Currently, nearly all the OTA signaling signals are design for intra-cell communication, e.g., the communication between the mobile device and its serving base station and the signaling between base stations is transmitted through wired backhaul.

*2) Inter-cell OTA Signaling*

The inter-cell OTA signaling refers to the signaling between the mobile device and its non-serving base station (neighboring base station) transmitted over the wireless channel. As discussed, inter-cell OTA signaling may be a better solution to deliver small amount of information between the mobile device and its non-serving base station in some applications, e.g., in femtocell networks. Since the femtocell network uses a third-party IP backhaul, the delay of this kind of backhaul is large and unpredictable, which may not be suitable for delivering some fast signaling with small amount of information. In femtocell networks, if the signaling goes directly between the mobile device and its non-serving base station over the wireless channel instead of first going between the mobile device and its serving base station over the wireless channel and then going between its serving base station and its non-serving base station over the backhaul, the signaling would be the fastest which satisfies the timely requirement for inter-cell coordination.

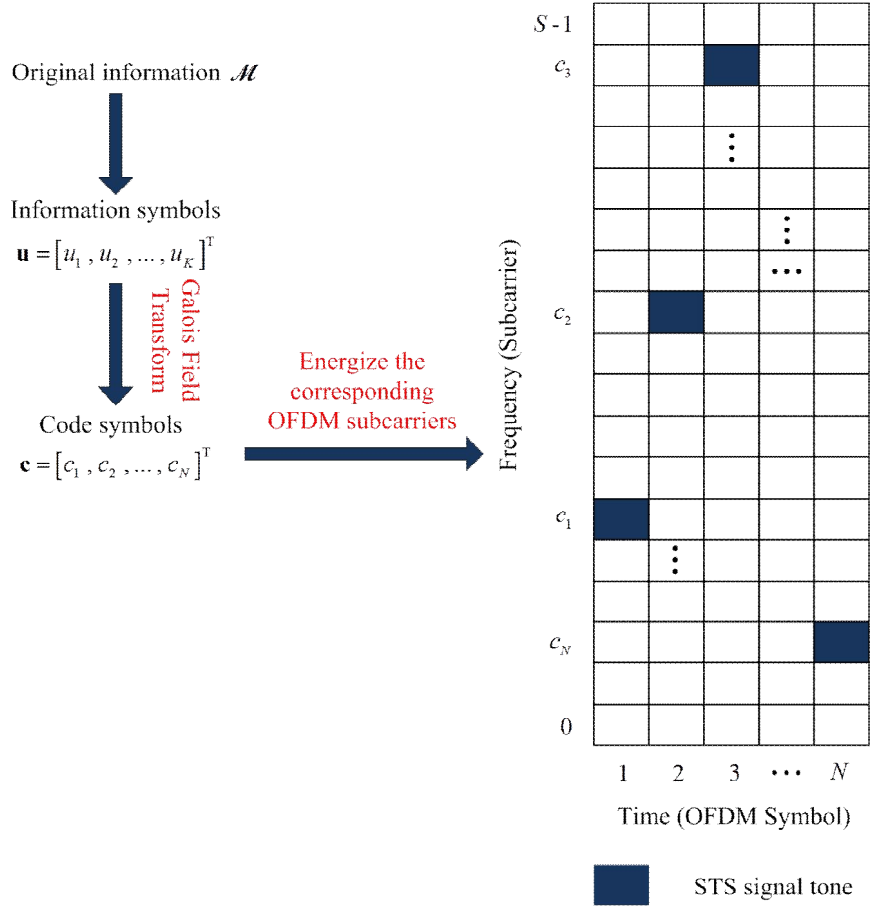

Figure 4. Procedure of STS processing at a transmitter.

## V. Proposed OTA Signature-based Signaling

In this section, we propose a new OTA signature-based signaling that can be used for inter-cell signaling in OFDM systems.

In our design, all the energy in an OFDM symbol is transmitted on a single OFDM subcarrier. No energy is transmitted on any other subcarriers of the current OFDM symbol and no information is modulated onto the energized subcarrier (i.e. neither amplitude nor phase is modulated). The energized subcarrier is referred to as the single-tone signaling (STS) signal tone. It is the index of the energized subcarrier that contains the information. That is, which subcarrier of this OFDM symbol is energized depends on the information that the STS signal carries.

The original information $\mathcal{M}$ is firstly represented by $K$ $(K \geq 1)$ information symbols, $\mathbf{u} = [u_1, u_2, \ldots, u_K]^T$, in the basis of $S$, or more precisely,

$$\mathcal{M}(S) = u_K S^{K-1} + u_{K-1} S^{K-2} + \ldots + u_2 S + u_1, \quad (1)$$

where $0 \leq u_k \leq S-1$, $1 \leq k \leq K$. Assume $S$ (consecutive or non-consecutive) subcarriers, numbered from 0 to $S-1$, are allowed for STS transmission. An STS signal can then be constructed by mapping the $K$ information symbols, $u_1, u_2, \ldots, u_K$, onto the $S \times K$ OFDM resource grid. The $u_k$ th subcarrier of the $k$ th OFDM symbol $(1 \leq k \leq K)$ is energized to form an STS signal tone. Therefore an STS signal consists of a sequence of $K$ tones. Hence $K$ OFDM symbols are needed to transmit message $\mathcal{M}$. Although the indistinctiveness nature of the signal tones prevents the STS signal tones from interfering between different STS signals, simultaneous transmissions of STS signals from multiple transmitters may confuse the receiver. To make the STS signal possess an ambiguity resolution capability and a certain degree of error protection against tone detection errors, we devise a special transform in a Galois field, $GF(S)$, as follows:

$$\mathbf{c} = \mathbf{G} \begin{bmatrix} 0 & \mathbf{u}^T & 0 & \cdots & 0 \end{bmatrix}^T, \quad (2)$$

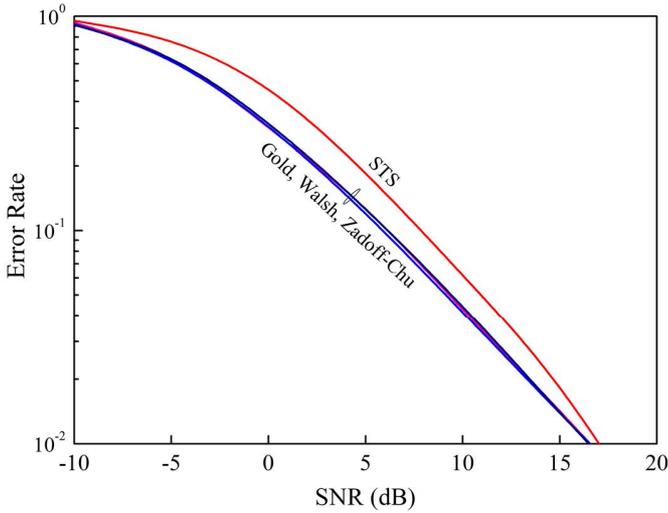

Figure 5. Single-signal detection performance of signature-based signaling. (One transmit antenna, one receive antenna, EPA channel model [13], fading speed is 3kmph)

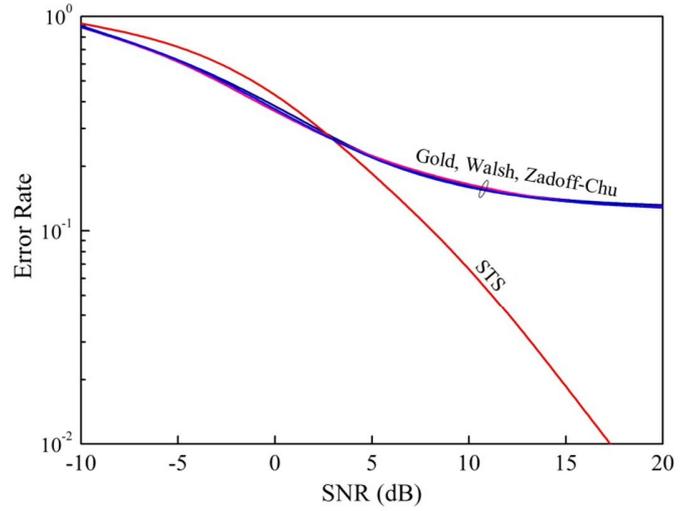

Figure 6. Two-signal detection performance of signature-based signaling. (One transmit antenna, one receive antenna, EPA channel model [13], fading speed is 3kmph)

where $\mathbf{c}$ is an $N \times 1$ vector, and

$$\mathbf{G} = \begin{bmatrix} 1 & 1 & \cdots & 1 \\ 1 & \alpha^{\frac{S-1}{N}} & \cdots & \alpha^{\frac{S-1}{N}(N-1)} \\ \vdots & \vdots & \ddots & \vdots \\ 1 & \alpha^{\frac{S-1}{N}(N-1)} & \cdots & \alpha^{\frac{S-1}{N}(N-1)(N-1)} \end{bmatrix}, \quad (3)$$

is an $N \times N$ matrix. Here $\alpha$ is a primitive number in $GF(S)$, $0 \leq c_n \leq S-1$ for $1 \leq n \leq N$, and $N \geq K$. One can consider (2) as a special non-binary coding in $GF(S)$ that transforms a sequence of the subcarrier indices (non-binary) of the STS signal tones, i.e., the original non-binary information symbols $\mathbf{u} = \begin{bmatrix} u_1 & u_2 & \ldots & u_K \end{bmatrix}^T$, into a code word $\mathbf{c} = \begin{bmatrix} c_1 & c_2 & \ldots & c_N \end{bmatrix}^T$ with code rate $(N,K)$, where $c_n$, $1 \leq n \leq N$ is the $n$th code symbols.

The procedure of STS processing at a transmitter is shown in Figure 4.

The proposed STS possesses many highly desirable and crucial properties for inter-cell OTA signaling, such as freedom from the inter-signal interference, immunity to synchronization error, and lack of the need for dedicated resources. Besides, the proposed STS signal provides a way of estimating inter-cell channel. The detection performance of the proposed signaling will be shown in the next section.

## VI. DETECTION PERFORMANCE

In this section, we compare the detection performance of the proposed STS described in Section V and several commonly used OTA signature-based signaling schemes based on Walsh sequence, Gold sequence, and Zadoff-Chu sequence. In the simulations, we assume time and frequency synchronization has been completed. For the proposed STS scheme, the number of subcarriers used for STS transmission is 71, the code rate of STS is $K/N = 1/14$; for the Walsh sequence based OTA signaling, the length of Walsh sequence is 1024; for Gold sequence based OTA signaling, the length of Gold sequence is 1023; for Zadoff-Chu sequence based OTA signaling, the length of Zadoff-Chu sequence is 1021. Each signaling scheme occupies a continuous time-frequency resources which composed of about 73 subcarriers across 14 OFDM symbols (a subframe in LTE).

Figure 5 shows the single-signal detection performance. An error is defined as the event that a signaling message is detected whereas it is not transmitted by any transmitter. The signal-to-noise ratio (SNR) is defined as the OFDM symbol SNR.

From Figure 5, we can observe that: (1) The OTA signature-based signaling that use Walsh sequence, Gold sequence, and Zadoff-Chu sequence all have similar detection performance. The reason is that all these signaling schemes almost have the same processing gain. (2) In the single-signal scenario, the detection performance of the proposed STS is ~2 dB worse than that of the other schemes. This is because there is no interference among different sequences in single-signal scenario. In the following, we will compare the performance in multi-signal scenario. For simplicity, we compare the detection performance in two-signal scenario. Figure 6 shows the two-signal detection performance where we assume that the receiver knows the number of transmitted signaling messages.

From Figure 6, we can observe that: (1) Like in the single-signal scenario, the OTA signature-based signaling that use Walsh sequence, Gold sequence, and Zadoff-Chu sequence all have similar detection performance in two-signal scenario but they have error floor due to inter-signal interference. (2) In two-signal scenario, the detection performance of the proposed STS is much better than that of the other schemes in median and high SNR regions though ~2 dB worse in low SNR region which is not in the operating range.

Compare Figure 5 and Figure 6, we can observe that: (1) The

OTA signature-based signaling that use Walsh sequence, Gold sequence, and Zadoff-Chu sequence have poorer detection performance in two-signal scenario than that in single-signal scenario due to interference between different signals, which is especially damage in high SNR region. (2) The detection performance of the proposed STS is almost the same both in single-signal and two-signal scenarios. This means that the proposed STS has better multi-signal detection performance which is a natural result since the different STS signals are transmitted on different subcarriers thus they have strong inter-signal interference resistance.

## VII. CONCLUSION

With the development of cellular networks, OTA signaling will play a more and more important role in future wireless networks. Various OTA signaling schemes are needed to facilitate the operation of many advanced technologies. In this article, we first give a brief introduction of OTA signaling in cellular networks taking LTE as an example. Then we describe the challenges and requirements in designing OTA signaling in detail. We give two important classifications of OTA signaling and compare their properties and applications. For inter-cell OTA signaling, we proposed a signature-based signaling named STS that has much better detection performance than commonly used schemes in multi-signal scenario.